\documentclass{aa}
\usepackage{graphics}
\newcommand\jcd{Christensen-Dalsgaard}
\newlength{\figwidth}
\newcommand\xib{\pmb{\xi}}
\newcommand\magb{{\bf B}}
\newcommand\vb{{\bf v}}
\newcommand\ope{{\cal L}}
\newcommand\calm{{\cal M}}
\newcommand\caln{{\cal N}}
\newcommand\calb{{\cal B}}
\newcommand\be{\begin{equation}}
\newcommand\ee{\end{equation}}
\def\pmb#1{\setbox0=\hbox{$#1$}\kern-0.015em\copy0\kern-\wd0%
\kern0.03em\copy0\kern-\wd0\kern-0.015em\raise0.03em\box0}

\def\note #1]{\box1{\bf #1]}}

\begin{document}
\thesaurus{06.01.2; 06.09.1; 06.13.1; 06.15.1}
\title{The Sun's acoustic asphericity and magnetic fields in the solar
convection zone}
\titlerunning{Magnetic field in solar convection zone}
\author{H. M. Antia\inst{1}\and S. M. Chitre\inst{1}\and M. J. Thompson\inst{2}}
\authorrunning{Antia et al.}
\offprints{H. M. Antia}
\institute{Tata Institute of Fundamental Research,
Homi Bhabha Road, Mumbai 400 005, India \and
Astronomy Unit, Queen Mary and Westfield College, Mile End Road,
London E1 4NS, U.~K.}
\date{Received }
\maketitle

\begin{abstract}
The observed splittings of solar oscillation frequencies can be employed
to separate the effects of internal solar rotation and to estimate
the contribution from a large-scale magnetic field or any
latitude-dependent thermal perturbation inside the Sun.
The surface distortion estimated from the rotation rate in
the solar interior is found to be in good agreement with the
observed oblateness at solar surface. After subtracting out the
estimated contribution from rotation, there is some residual signal
in the even splitting coefficients, which may be explained by
a magnetic field of approximately 20 kG strength located at a depth of 30000
km below the surface or an equivalent aspherical thermal perturbation.
An upper limit of 300 kG is derived
for a toroidal field near the base of the convection zone.

\end{abstract}

\section{Introduction}

Rotational splittings of solar oscillation frequencies have been successfully
utilized to infer the rotation rate in the solar interior.
To first order, rotation affects only the splitting coefficients which
represent odd terms in the azimuthal order $m$ of the global resonant modes.
The even terms in these splitting coefficients, which reflect the
Sun's effective acoustic asphericity, can arise
from second order
effects contributed both by the rotation and magnetic field as also
from latitudinal temperature variations.
Since the rotation rate
can be inferred using the odd splitting coefficients, the inferred
profile can be used to estimate the second order effects. These can
then be subtracted from the observed even coefficients to
estimate the magnetic field strength (Gough \& Thompson 1990) or other
latitudinal variations in sound propagation speed.
The distortion introduced by rotation can be compared with the measured
oblateness at the solar surface.

The even coefficients of splittings are fairly small, and
no definitive results have so far been obtained regarding
the magnetic field strength in the solar
interior. With the good quality data now becoming available from
GONG (Global Oscillation Network Group) and MDI (Michelson Doppler
Imager) projects, it is desirable to investigate the possibility
of inferring the strength of magnetic field in solar interior.
There should also be some shift in the mean frequency for each
$n,\ell$ multiplet due to second order effects from rotation
and magnetic field, which can also be estimated. It is difficult
to measure this frequency shift from observed data as it is hard
to separate it from the effects of other uncertainties in the spherical
structure of the Sun.
Nevertheless, these frequency shifts can affect
the helioseismic inferences and it would be interesting to estimate
their effect.

\section{The technique}

The frequencies of solar oscillations can be expressed in terms of
the splitting coefficients:
\be
\nu_{n,\ell,m}=\nu_{n,\ell}+\sum_{j=1}^{J_{\rm max}} a_j^{n,\ell}
{\cal P}_j^{\ell}(m),\qquad\qquad (J_{\rm max}\le 2\ell)
\label{split}
\ee
where ${\cal P}_j^\ell(m)$ are orthogonal polynomials of degree $j$ in $m$
(Ritzwoller \& Lavely 1991; Schou, \jcd\ \& Thompson~1994).
The odd coefficients $a_1,a_3,a_5,\ldots$
can be used to infer the rotation rate in the solar interior, while the
even coefficients arise basically from second order effects due to rotation and
magnetic field. Since forces due to rotation or magnetic field in the
solar interior are smaller by about 5 orders of magnitude
as compared to gravitational forces, it is
possible to apply a perturbative treatment to calculate their
contribution to frequency splittings. In this approach, we estimate
the effects of rotation and magnetic field on the frequencies but
without explicitly constructing a model of a rotating, magnetic star.

We adopt the formulation due to Gough \& Thompson (1990), with the
difference that
we include perturbation in the gravitational potential and also
assume differential rotation in the interior,
though the symmetry axis of magnetic field is taken to coincide
with rotation axis.

In an inertial frame the oscillation equations can be formally written as
\be
{\cal L}\pmb{\xi}+\rho\omega^2\pmb{\xi}=\omega{\cal M}\pmb{\xi}
+{\cal N}\pmb{\xi}+{\cal B}\pmb{\xi},
\label{osc}
\ee
where
\begin{eqnarray}
{\cal L}\pmb{\xi}&=&\nabla(\rho c_s^2\nabla\cdot\xib+\xib\cdot\nabla p)
-(\nabla\cdot\xib+\xib\cdot\nabla\ln\rho)\nabla p\nonumber\\
& &\qquad-\rho G\nabla\left(\int {\nabla\cdot(\rho\xib)\over|{\bf r}-{\bf r}'|}
d^3{\bf r}'\right),
\\
{\cal M}\pmb{\xi}&=& -2i\rho {\bf v}\cdot\nabla\pmb{\xi},\\
{\cal N}\pmb{\xi}&=&-\rho\pmb{\xi}\cdot\nabla({\bf v}\cdot\nabla{\bf v})
+\rho({\bf v}\cdot\nabla)^2\pmb{\xi},\\
{\cal B}&=&-{1\over4\pi}\left(
{\nabla\cdot(\rho\pmb{\xi})\over\rho}(\nabla\times \magb)
\times\magb+(\nabla\times\magb_1)\times\magb\right.\nonumber\\
& &\left.\phantom{\nabla\cdot(\rho\pmb{\xi})\over\rho}+(\nabla\times\magb)
\times \magb_1\right).
\label{terms}
\end{eqnarray}
Here $\magb_1=\nabla\times(\xib\times\magb)$ is the linearized Eulerian
perturbation to magnetic field, $\magb$,
$\xib$ is the displacement eigenfunction,
$\vb=\pmb{\Omega}\times{\bf r}$ is the velocity due to rotation,
and  $p,\rho,
c_s$ are respectively, pressure, density and sound speed in the equilibrium
state.

In the presence of rotation and magnetic field the equilibrium
state will naturally undergo a distortion that needs to
be included in the calculations.
To account for this deformation we consider a transformation
to map each point $\bf r$ in the distorted star to  a point $\bf x$ in the
spherical volume occupied by the undistorted star by a transformation
\be
x=(1+h_\Omega({\bf r})+h_B({\bf r}))r,\label{dist}
\ee
where the functions $h_\Omega({\bf r})$ and $h_B({\bf r})$ which
depend on the rotation and magnetic field respectively,
are to be determined by
solving the equations for equilibrium in a distorted star
(Gough \& Thompson 1990).
This will give us the perturbation to a nonrotating spherically
symmetric solar model and the extent of distortion at the surface
may be compared with observed values.
Here, $x$ is chosen so that $x=R$ can be regarded as the
distorted solar surface, where $R$ is the radial distance of the outermost
layer included in the solar model. Similarly,
various equilibrium quantities are also expressed in the form
\be
\rho(r)=\rho_0(x)+\rho_\Omega({\bf x})+\rho_B({\bf x}).
\label{rhodist}
\ee
In all these expansions higher order terms have been neglected.

We consider the terms on the right hand side of Eq.~\ref{osc},
as perturbations
to basic equations for linear adiabatic oscillations for non-magnetic
and non-rotating star. Rotation introduces a first order
perturbation through $\calm$ which gives the odd splitting
coefficients, while magnetic field can only give rise to
even  terms in $m$ and contributes to the even splitting coefficients.
The distortion from a spherically symmetric equilibrium state
also introduces even order terms.
The relative magnitude of contributions from rotation
and magnetic field will, of course,
depend on the rotation rate and magnetic field strength.
For the solar case we know that odd splitting coefficients
arising from the first order effect of rotation are much larger than the
even coefficients and we therefore expect the magnetic field to make
a comparatively smaller contribution. We must therefore include
the effect of
rotation to second order, while magnetic field and distortion effects
need be retained only to first non-vanishing terms. The first order
perturbation arising in frequencies on account of rotation
also introduces
a perturbation to eigenfunctions which will give a second order
contribution. We can formally express the frequency and eigenfunction as
\be
\omega=\omega_0+\omega_1+\omega_2,\qquad \xib=\xib_0+\xib_1.
\label{pert}
\ee
Retaining terms to second order, we get
\begin{eqnarray}
&\ope_0(\xib_0+\xib_1)+\ope_\Omega\xib_0+\ope_B\xib_0+
\rho_0(\omega_0^2+2\omega_0\omega_1)(\xib_0+\xib_1)\nonumber\\
&\qquad+\rho_0(\omega_1^2+2\omega_0\omega_2)\xib_0+
\rho_\Omega\omega_0^2\xib_0+\rho_B\omega_0^2\xib_0\nonumber\\
&\qquad =\omega_0\calm(\xib_0+\xib_1)+\omega_1\calm\xib_0+
\caln\xib_0+\calb\xib_0.
\label{pertii}
\end{eqnarray}
Here, $\ope_\Omega$ and $\ope_B$ are the perturbations 
to $\ope$ arising from distortion of equilibrium state due to rotation
and magnetic field respectively.
Taking the scalar product with $\xib_0^*$ and integrating over the
entire volume, we recover
\begin{eqnarray}
& &2\omega_0\langle \rho_0\xib_0^*\xib_0\rangle \omega_2=
\langle\xib_0^*(\caln-\ope_\Omega-\rho_\Omega\omega_0^2)\xib_0\rangle\nonumber\\
& &\qquad+\langle\xib_0^*(\calb-\ope_B-\rho_B\omega_0^2)\xib_0\rangle
-\omega_1^2\langle\rho_0\xib_0^*\xib_0\rangle\nonumber\\
& &\qquad-2\omega_0\omega_1\langle\rho_0\xib_0^*\xib_1\rangle
+\omega_1\langle\xib_0^*\calm\xib_0\rangle
+\omega_0\langle\xib_0^*\calm\xib_1\rangle,
\label{dnu}
\end{eqnarray}
where the angular brackets denote
\be
\langle f(x,\theta,\phi)\rangle=\int_{x<R} f(x,\theta,\phi) x^2\sin\theta
\;dx\;d\theta\;d\phi
\label{scprod}
\ee
The first order correction to frequency is given by
\be
\omega_1={\langle\xib_0^*\calm\xib_0\rangle\over
2\langle\rho_0\xib_0^*\xib_0\rangle},
\label{dnurot}
\ee
while perturbation to the eigenfunction may be calculated using
\be
\ope\xib_1+\rho_0\omega_0^2\xib_1=-2\rho_0\omega_0\omega_1\xib_0
+\omega_0\calm\xib_0.
\label{egn}
\ee

The observed odd splitting coefficients can be used to infer the
rotation rate inside the Sun (Thompson et al.~1996; Schou et al.~1998). We
approximate this rotation rate using the first three terms in
the expansion of the angular velocity,
\be
\Omega(r,\theta)=\Omega_0(r)+\Omega_2(r)\cos^2\theta+
\Omega_4(r)\cos^4\theta,
\label{rot}
\ee
where $\theta$ is the colatitude. This rotation rate is then used to
compute the second order rotational contribution to frequency splitting, which
may be subtracted from the observed splittings to obtain the residual
which may be due to magnetic field, any other velocity field or
asphericity in solar structure.

In the present analysis we use only the toroidal magnetic field,
taken to be of the form,
\be
{\bf B}=\left[0,0,a(r){dP_k\over d\theta}{(\cos\theta)\atop
\phantom{d\theta}}\right],
\label{mag}
\ee
with the axis of symmetry coinciding with the rotation axis.
Here $P_k(x)$ is the Legendre polynomial of degree $k$. 
The Lorentz force due to a field of this form can be written as
\be
{\bf F}=\rho(r)\sum_{\lambda=0}^k\left[f_{r\lambda}(r)
P_{2\lambda}(\cos\theta),f_{\theta\lambda}(r){dP_{2\lambda}\over
d\theta},0\right].
\label{lorfor}
\ee
Each of this term can be treated separately and the results can be
combined to yield the net effect.

We calculate the second order frequency shift due to rotation
and magnetic field for each value of $m$ and then use Eq.~\ref{split}
to obtain the corresponding splitting coefficients. These can
then be compared with observed coefficients from GONG
(Hill et al.~1996) or MDI (Rhodes et al.~1997) data.
To evaluate the angular integrals we use the following recursion
relations
\begin{eqnarray}
\cos\theta Y_\ell^m&=&C_\ell^m Y_{\ell+1}^m+C_{\ell-1}^mY_{\ell-1}^m,\\
\sin\theta {\partial Y_\ell^m\over \partial\theta}&=&
\ell C_\ell^m Y_{\ell+1}^m-(\ell+1)C_{\ell-1}^mY_{\ell-1}^m,
\label{ylm}
\end{eqnarray}
where
\be
C_\ell^m=\sqrt{(\ell+1+m)(\ell+1-m)\over (2\ell+1)(2\ell+3)}.
\label{clm}
\ee
Since we have used only the first two terms in the expansion of
rotation rate as a function of latitude, we restrict to
calculation of the splitting coefficients $a_2$ and $a_4$ in this
work.

\section{Results}

We use the rotation rate inferred from the GONG data for the
months 4--14 (Antia, Basu \& Chitre 1998) to estimate the second
order frequency shift and the corresponding splitting coefficients
$a_2$ and $a_4$, as outlined in the previous section.
We incorporate all the second-order contributions arising from rotation, including
those from the distortion of equilibrium state and the
perturbation to the eigenfunctions.
Although there may be some variation in rotation rate with time, the
estimated variation is very small and its effect on the inferred splitting
coefficient would be much smaller than the errors in observed values.

\subsection{Shift in the mean frequency}

In principle, the shift in the mean frequency arising from
second order effects of rotation can
be calculated with the help of the prescription outlined in the previous
section, by taking the
spherically symmetric component of the perturbing force ($\lambda=0$
term in Eq.~\ref{lorfor}).
However, this will also change the mass, radius and luminosity of the
solar model. The change may be smaller than the errors in
observed radius or luminosity, but it may tend to give a different estimate
for modified frequency compared to what will be obtained if
the observed constraints on mass, radius and luminosity were
to be exactly applied.
Hence, for obtaining a consistent estimate of the effect of
distortion, we construct a spherically symmetric solar model
with correct mass, radius and luminosity by modifying the effective
acceleration due to gravity, $g$ to account for the spherically
symmetric component of forces due to rotation. The difference in frequency
of this model in relation to a standard, non-rotating model
would give the frequency shift due to distortion. All the other
second-order rotational
terms are added to this shift, to obtain the
total shift in frequency due to rotation which is displayed in Fig.~1.
This figure includes all modes with $0.5<\nu<4.5$ mHz and $\ell\le250$.
The corrections to mean frequencies due to general relativistic effects as
discussed towards the end of this subsection, are also shown in the figure.

\begin{figure}
\resizebox{\figwidth}{!}{\includegraphics{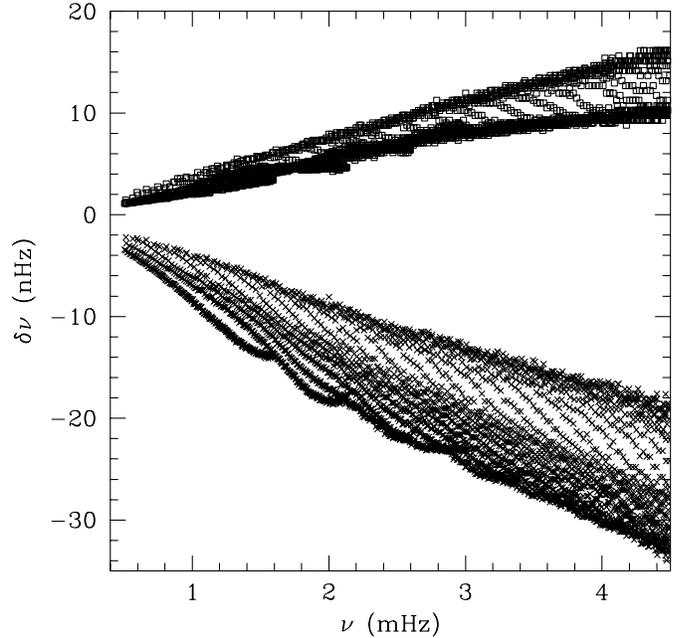}}
\caption{
The shift in mean multiplet frequencies
due to second order effects from
rotation is shown by the crosses, while the squares (points with
positive $\delta\nu$) show the frequency
shift due to general relativistic effects.}
\end{figure}

This relative frequency shift, which is less
than $10^{-5}$, is nonetheless comparable to 
the estimated errors in the observed
frequencies and the correction should, in principle, be applied while
doing inversions (e.g. Gough et al.~1996) for the Sun's spherical structure.
In order to estimate the error introduced
by neglect of this effect, we can carry out an inversion for sound speed
and density in the solar interior using this frequency shift due to
rotation as the frequency difference and the results are shown in
Fig.~2.
The inversions are performed using a regularized least squares
inversion technique (Antia 1996).
The resulting $\delta c_s^2/c_s^2$ and $\delta\rho/\rho$ are almost an
order of magnitude
less than the estimated errors in inversions.

As an aside, we note that the internal rotation rate from 
Antia et al.~(1998) adopted in our study was obtained assuming 
a spherically symmetric background state for the Sun, as is usual for
inversions for the solar rotation.  We realise that both the
mean frequencies of solar oscillations and the rotational
splittings will be modified by departures in the equilibrium solar
model from spherical symmetry, as discussed in this paper. 
In order to estimate the resulting
shift in rotational splittings we would need to calculate the third
order terms in perturbation expansion of Gough \& Thompson~(1990).
We have not included these terms in our analysis,  but we expect that their
contribution would have the same relative magnitude of $10^{-5}$
as that found for the shift in mean frequencies. This is clearly,
much smaller
than the estimated errors in splitting coefficients in current
helioseismic data sets. Therefore we do not expect the rotational
splittings and hence the inverted rotation rate to be significantly
affected by this higher order effect.

\begin{figure}
\resizebox{\figwidth}{!}{\includegraphics{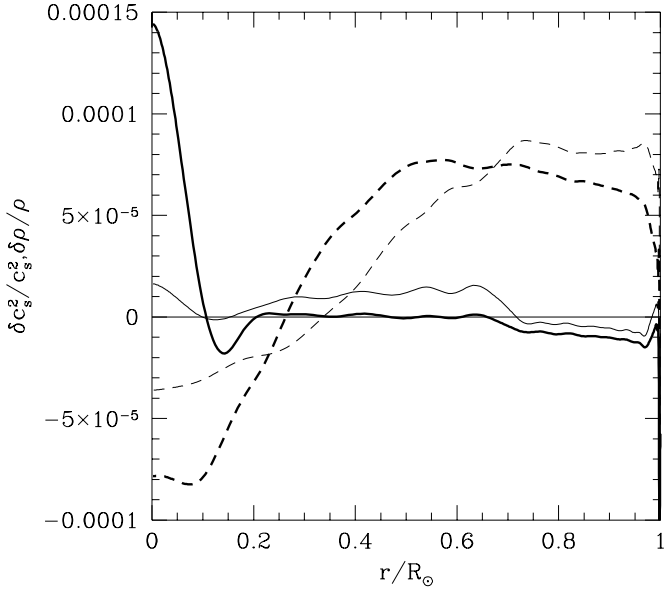}}
\caption{
The correction to sound speed and density as inferred
from helioseismic inversions, arising from
the frequency shifts shown in Fig.~1. The thick continuous and short-dashed
lines show, respectively, $\delta c_s^2/c_s^2$ and $\delta\rho/\rho$
due to frequency shifts from rotation (crosses in Fig.~1), while the
thin lines show the same arising from combined frequency shifts
due to rotation and general relativity.
}
\end{figure}

It may be noted that mean frequencies of f-modes
get diminished by up to 15 nHz on account of the effect of rotation.
Since rotation effectively reduces the acceleration due to gravity $g$,
this leads to a decrease in the frequencies of f-modes.
The relative change in
f-mode frequencies is shown in Fig.~3. If this effect is taken
into account the estimated solar radius using f-mode frequencies (Schou et
al.~1997, Antia 1998) would effectively be decreased by
about 4 km. This is again much less than the systematic errors in
estimated radius, though the decrease is larger than the statistical
errors (Tripathy \& Antia 1999).

\begin{figure}
\resizebox{\figwidth}{!}{\includegraphics{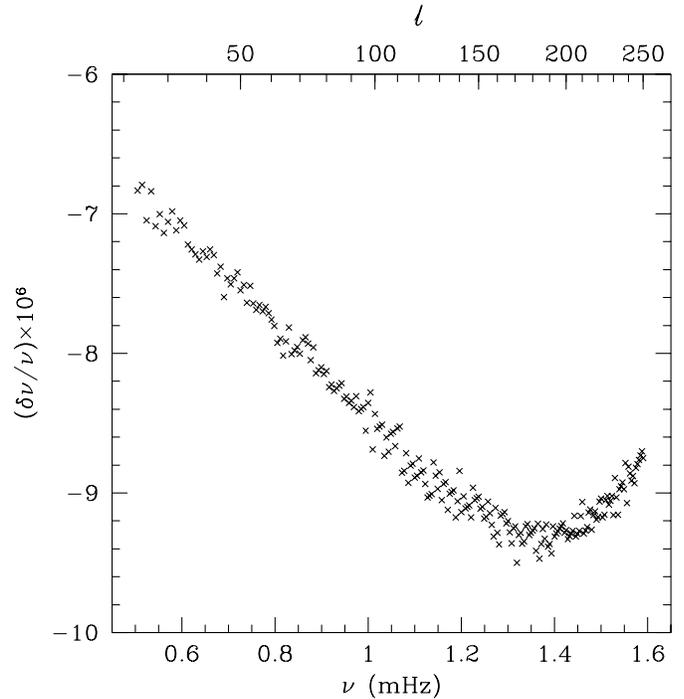}}
\caption{
The relative shift in mean multiplet f-mode frequencies due to rotation}
\end{figure}

It is interesting to note that apart from second order effects of
rotation, there would also be corrections to the frequencies arising
from general relativity. The relativistic effect can be measured
by $Gm(r)/(rc^2)$, where $G$ is the gravitational constant, $m(r)$
is the mass contained within spherical shell of radius $r$, and 
$c$ the speed of light. Fig.~4 shows this ratio in a solar model
and it can be seen that it is comparable to the ratio of centrifugal to
gravitational forces. It is possible to calculate a solar model
using Oppenheimer-Volkoff equation of relativistic stellar
structure instead of the standard equation of hydrostatic equilibrium:
\be
{dp\over dr}=-{G(\rho+p/c^2)(m+4\pi r^2p/c^2)\over
r^2(1-2Gm/rc^2)}.
\label{opvol}
\ee
It is clear that general relativistic effect would be of opposite
sign to that due to rotation, as rotation effectively reduces the
acceleration due to gravity, $g$, while the relativistic correction 
tends to increase
it. Thus there is a partial cancelation between the two effects.
It is possible to calculate the change in solar models due to
the relativistic effect, although a detailed calculation of
frequencies using relativistic stellar oscillations equations would
require considerable effort and is beyond the scope of the present work.
To a first approximation we may calculate the effect by using the
normal equations of stellar oscillations with gravity modified
according to Eq.~\ref{opvol}. Such a calculation shows that the effect of
relativity more or less cancels the frequency shift due to rotation
for low degree modes. The frequency shift due to general
relativity are also shown in Fig.~1. If this
frequency shift is added to the contribution arising from rotation
then the effect on helioseismic inversion is significantly reduced in
the solar core as can be seen from Fig.~2 (compare the thick and thin
lines).

\begin{figure}
\resizebox{\figwidth}{!}{\includegraphics{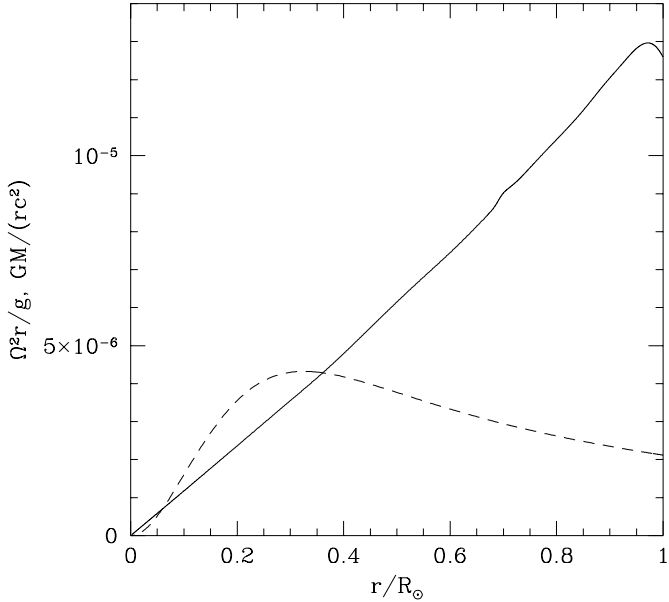}}
\caption{
The ratio of centrifugal force and the gravitational force
(continuous line) in a solar model is shown as a function of the
radial distance $r$. The dashed line shows the ratio $Gm(r)/(rc^2)$
which gives the effect due to general relativity.}
\end{figure}

\subsection{Oblateness due to rotation}

During the course of computing the splitting coefficients, it is necessary to
calculate the deformation induced by rotation as outlined by
Gough and Thompson~(1990). This deformation may be compared with
the observed oblateness at the solar surface.
The surface amplitudes of the $P_2(\cos\theta)$ and $P_4(\cos\theta)$
components of deformation are found to be $-5.84\times10^{-6}$ and
$6.2\times10^{-7}$ respectively, which are consistent with the estimates
obtained by Armstrong \& Kuhn~(1999). These can be compared with measured
values of $-(5.44\pm0.46)\times10^{-6}$ and $(1.48\pm0.58)\times10^{-6}$
respectively, from MDI measurement during 1997 (Kuhn et al.~1998).
Kuhn et al.~(1998) find a large temporal variation in the $P_4$
component, but it is not clear if the variation is statistically
significant.
It can be seen that the measured values of solar oblateness are
reasonably close to those expected from rotational distortion.
There may be some residual arising from other effects, like magnetic
field or other asphericities. The contribution from magnetic field is
indeed expected to vary with solar cycle and may account for the variation
in $P_4$ component, if the variation is in fact real.

It is also possible to estimate the global parameters for the Sun, like
angular momentum, rotational kinetic energy and gravitational
quadrupole and hexadecapole moments due to rotational distortion
(Pijpers 1998) and the results are summarized below:
\begin{eqnarray}
& &\hbox{Moment of Inertia, }  I =
7.11\times10^{53} {\rm gm\;cm}^2,\\
& &\hbox{Angular Momentum, }  H =
1.91\times10^{48} {\rm gm\;cm}^2\; {\rm s}^{-1},\\
& &\hbox{Kinetic Energy, }  T =
2.57\times10^{42} {\rm gm\;cm}^2\; {\rm s}^{-2},\\
& &\hbox{Quadrupole Moment, }  J_2 =
-2.18\times10^{-7},\\
& &\hbox{Hexadecapole Moment, }  J_4 = 
4.64\times10^{-9},
\end{eqnarray}
which are consistent with estimates of Pijpers (1998), who obtained his
estimates by working in terms of kernels for the various quantities. 
The value of
$J_2$ will yield a precession of the perihelion of planet
Mercury by about 0.03 arcsec/century, which is small enough to maintain
consistency of the general theory of relativity.

\subsection{Second order splitting due to rotation}

The contribution to splitting coefficients $a_2$ and $a_4$ due to rotation 
is shown in Fig.~5. This contribution needs to be subtracted from
the observed splitting coefficients for obtaining the residual contribution
which may arise from effects due to magnetic field, other velocity fields or
asphericity in solar structure.
Since the errors in individual
splitting coefficients are too large to give significant differences,
we average over 30 neighbouring modes in $w=\nu/(\ell+1/2)$ and the corresponding
results are shown in Fig.~6. There is reasonable agreement between
the GONG data for months 4--14 (23 August 1995 to 21 September 1996)
and MDI data for the first 360 days of its operation
(1 May 1996 to 25 April 1997). 
It is well known that the even splitting coefficients vary
with solar activity cycle (Libbrecht \& Woodard 1990; Dziembowski
et al.~1998; Howe, Komm \& Hill~1999) and there may not be agreement
between observations taken at different epochs. But in the present case
there is considerable overlap in period and the observations are near
the minimum phase of solar activity, when these coefficients are not
expected to vary significantly.

The difference between the observed
splitting coefficients and the estimated contribution from rotation is
significant for modes with turning points in the convection zone.
For modes penetrating more deeply, the errors are larger and the difference
is probably not significant.

\begin{figure*}
\hbox to \hsize{\resizebox{\figwidth}{!}{\includegraphics{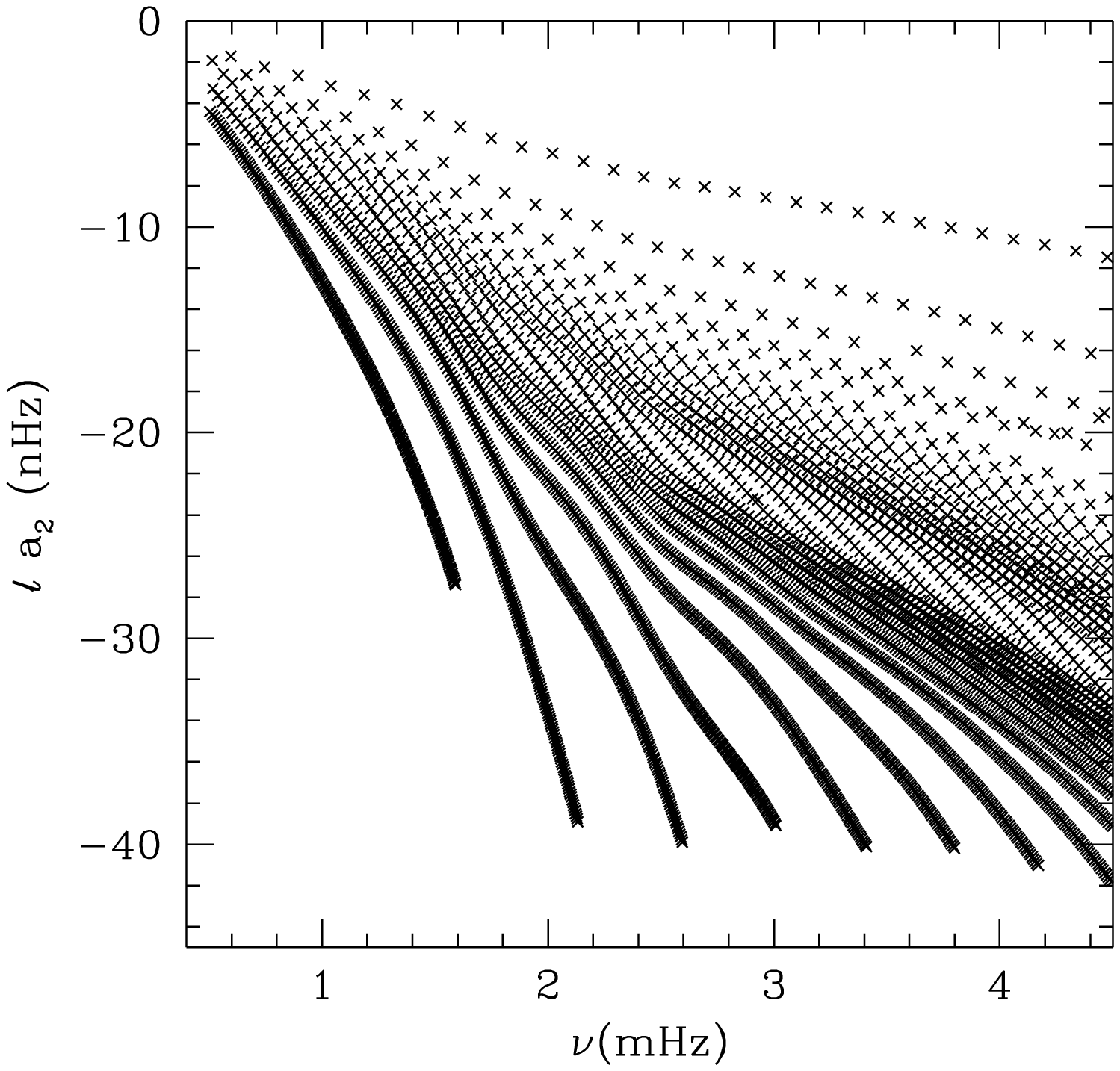}}
\hfil\resizebox{\figwidth}{!}{\includegraphics{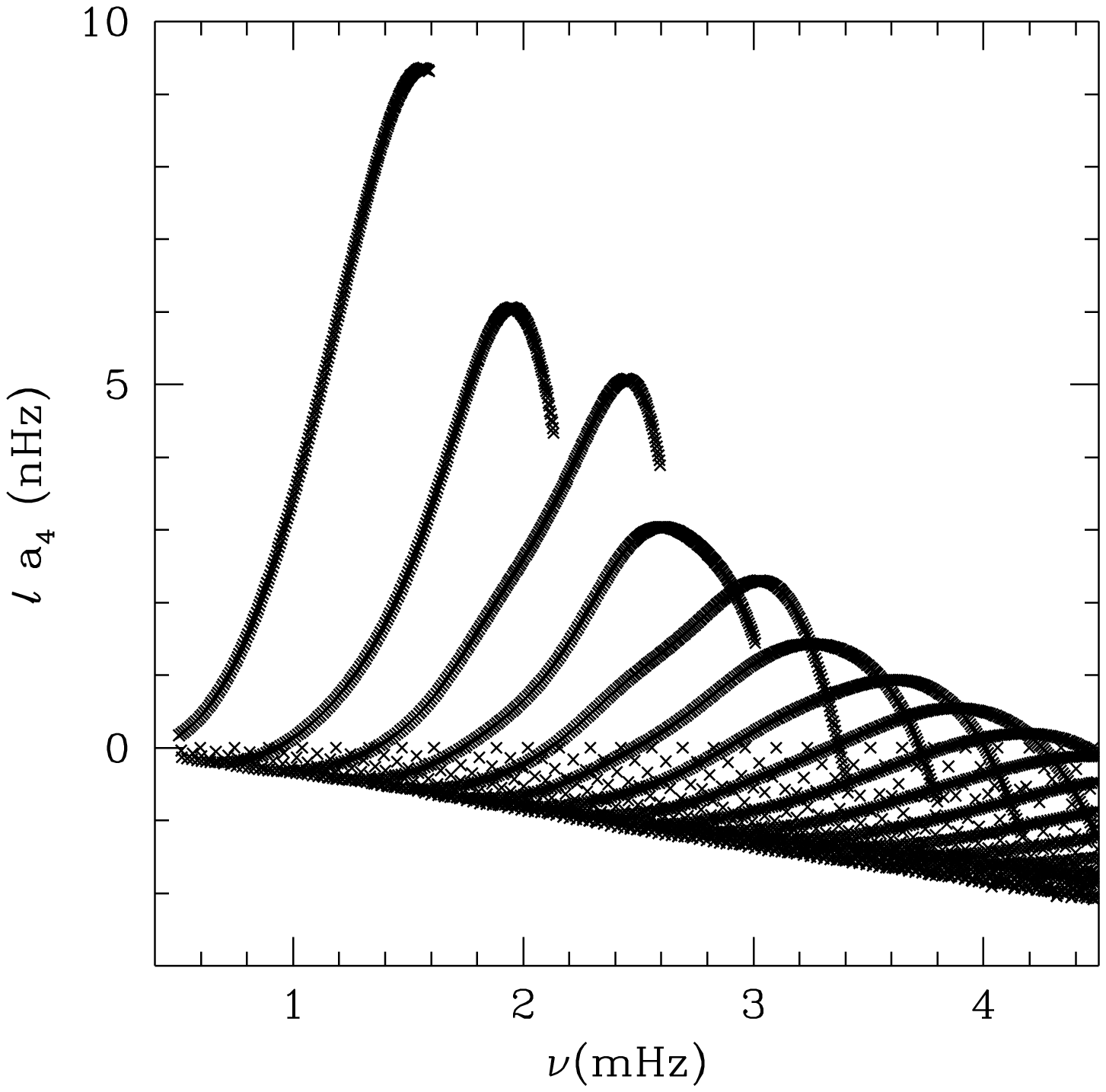}}}
\caption{
The splitting coefficients $a_2$ and $a_4$ from effects of
rotation.}
\end{figure*}

\begin{figure*}
\hbox to \hsize{\resizebox{\figwidth}{!}{\includegraphics{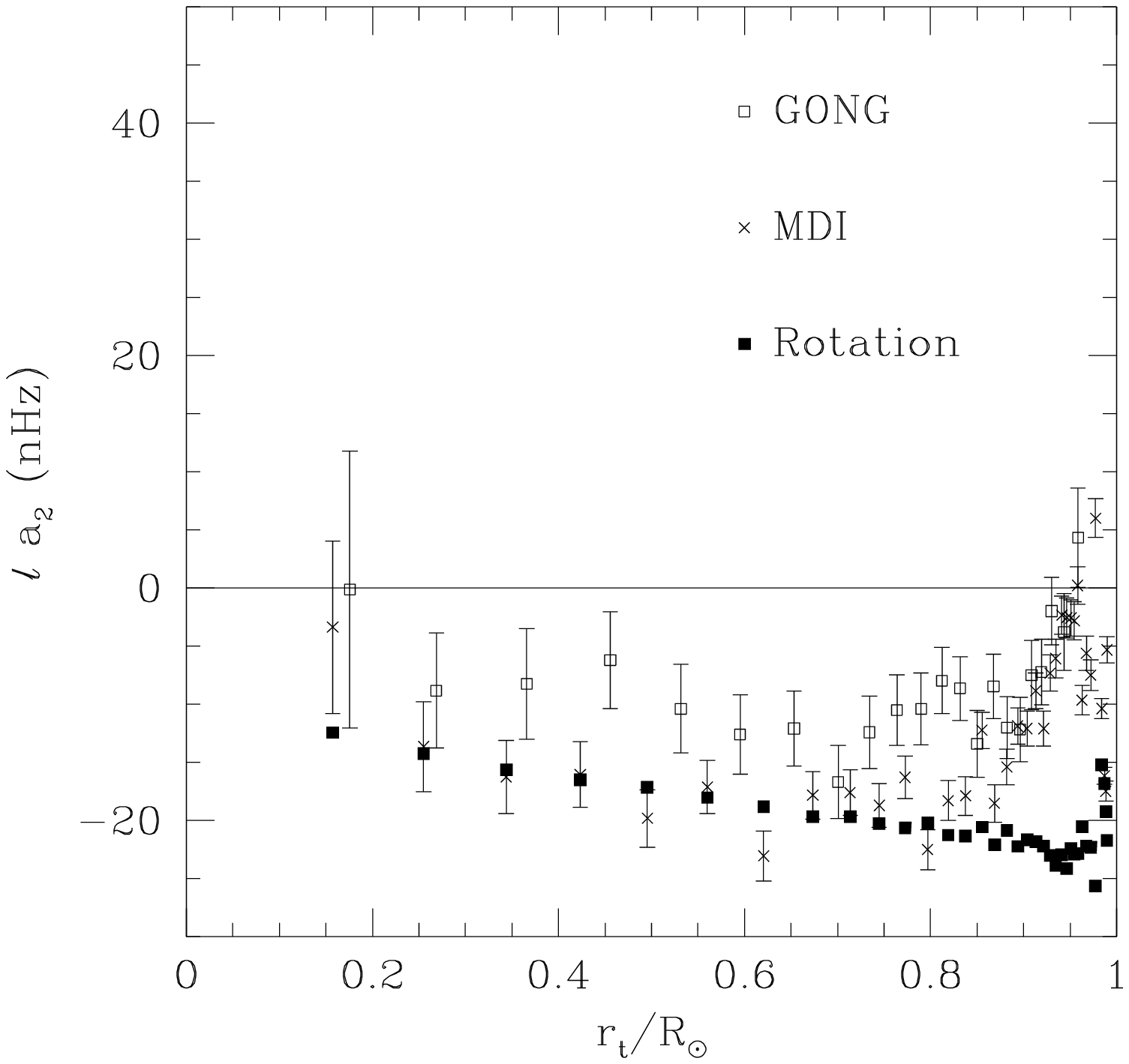}}
\hfil\resizebox{\figwidth}{!}{\includegraphics{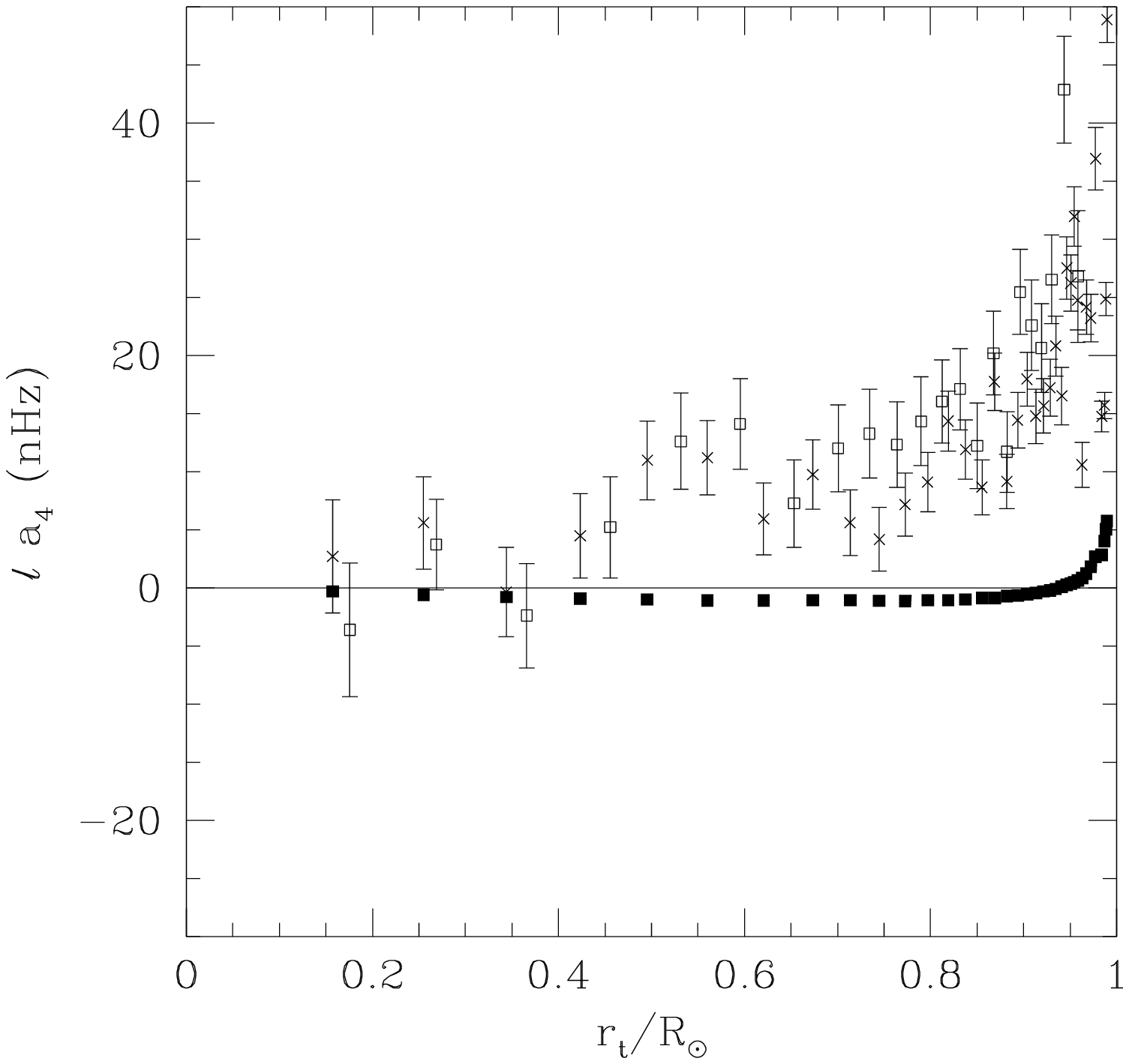}}}
\caption{
The splitting coefficients $a_2$ and $a_4$ plotted
as a function of the lower turning point for the modes. The
coefficients from GONG and MDI data are compared with the contribution
from rotation. Each point represents an average over 30 neighbouring modes.}
\end{figure*}

\subsection{Splitting due to magnetic field near the base of the convection zone}

There have been some suggestions that a significant
toroidal magnetic field may be
concentrated in a layer around the base of the
convection zone (Dziembowski \& Goode 1992).
We therefore first investigate splittings that
are expected
from such a field by assuming the magnetic field to be given by
Eq.~\ref{mag} with
\be
a(r)=\cases{\sqrt{8\pi p_0\beta_0}(1-({r-r_0\over d})^2)
&if $|r-r_0|\le d$\cr
0 &otherwise\cr}
\label{magtor}
\ee
where $p_0$ is the gas pressure, $\beta_0$ is a constant
giving the ratio of magnetic to gas pressure,
$r_0$ and $d$ are constants defining
the mean position and thickness of layer where the field is
concentrated. Fig.~7 shows the splitting coefficients resulting from a
toroidal magnetic field of this form concentrated at the base of the 
convection zone ($r_0 = 0.713 R_\odot$ and
$d = 0.02 R_\odot$). The splitting shown in this and subsequent
figures includes both the direct and distortion contributions as defined
by Gough and Thompson~(1990).

\begin{figure*}
\hbox to \hsize{\resizebox{\figwidth}{!}{\includegraphics{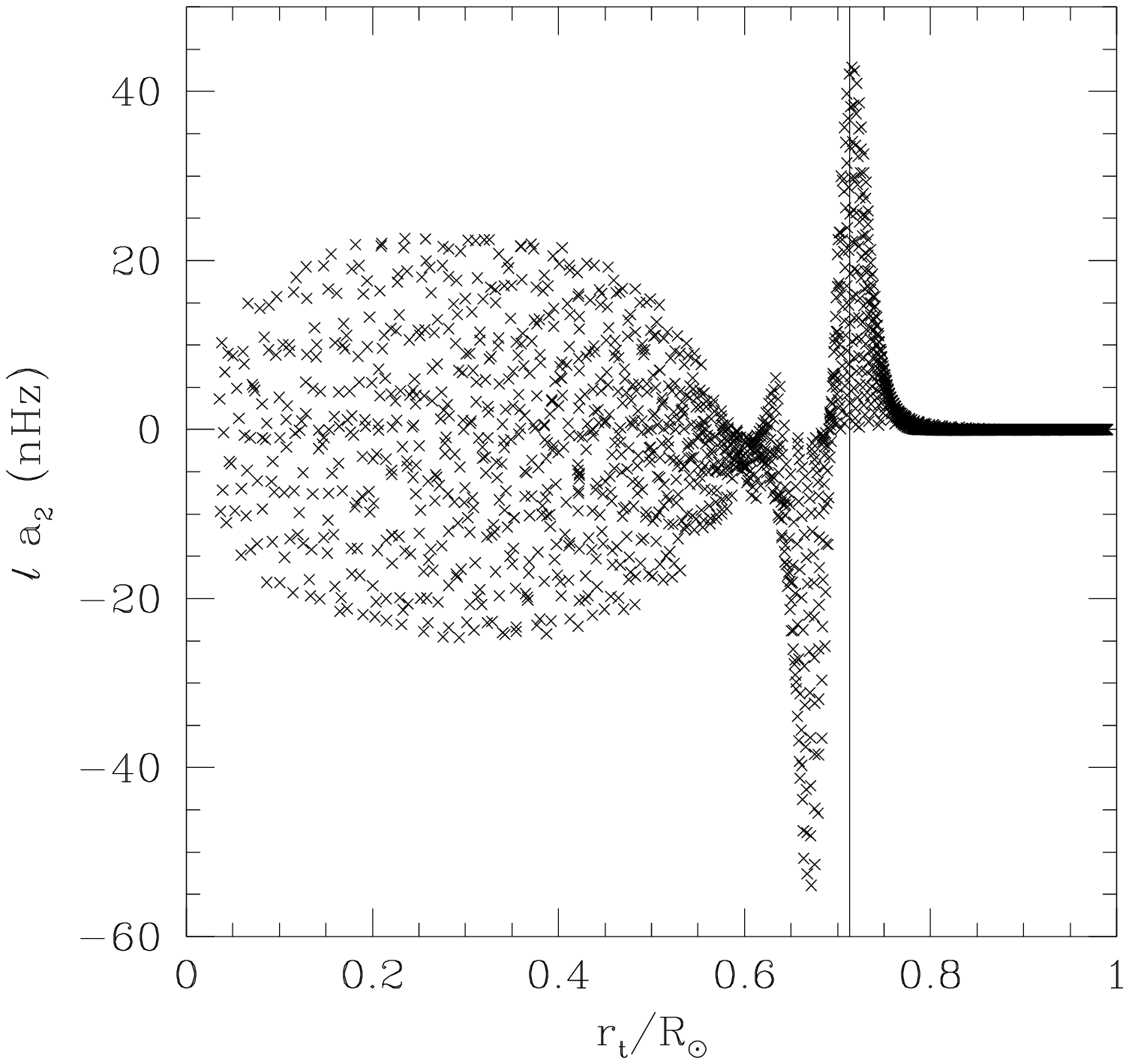}}
\hfil\resizebox{\figwidth}{!}{\includegraphics{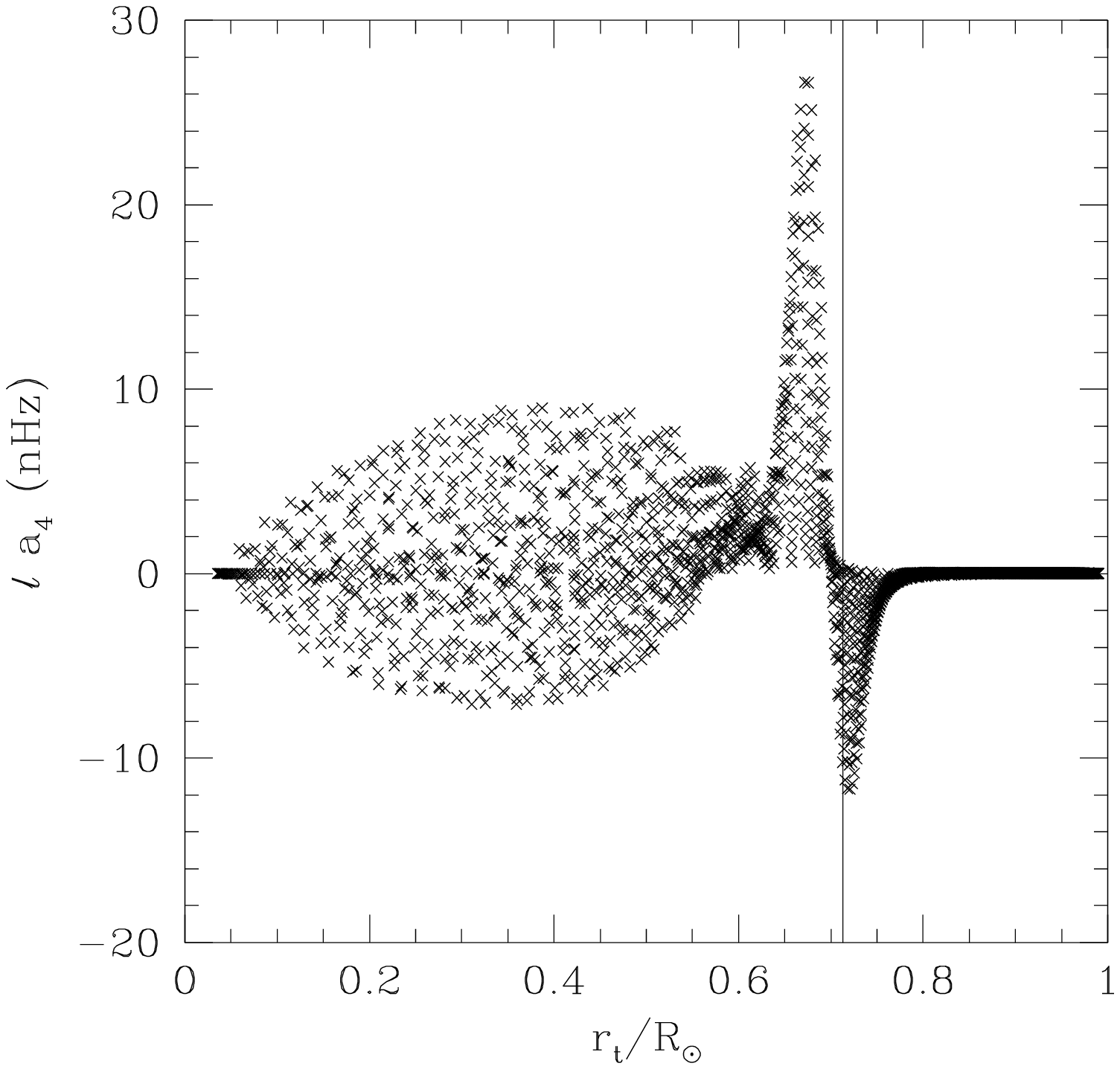}}}
\caption{
The splitting coefficients $a_2$ and $a_4$ from
a toroidal magnetic field concentrated near the base of the
convection zone, plotted
as a function of the lower turning point for the mode.
Magnetic field is given by Eqs.~(16,27) with $k=2$, $\beta_0=10^{-4}$,
$r_0=0.713R_\odot$ (shown by the vertical line in the figure)
and $d=0.02R_\odot$.}
\end{figure*}

The coefficients $a_2$ and $a_4$ from a toroidal magnetic field concentrated
near the base of the convection zone have a characteristic signature
for modes with turning point near the base of the convection zone;
it should be possible to detect such a signal in the observed
splittings if a strong enough magnetic field is indeed present in these
layers. 
The computed splittings, particularly for the deeply penetrating modes 
in Fig.~7 show a great spread, which is characteristic of the splittings
arising from a thin magnetic layer. We return below to the use that 
can potentially be made of this signature. In the present study, however,
we choose to average over neighbouring modes, as discussed in Section 3.3, 
which suppresses this
spread. Our rationale is that the errors in the real data
are too large for the spread to be visibly distinguished from noise in the 
measured splittings at present. Thus
we take
averages over neighbouring modes and compare the residual after removing
the contribution due to rotation with the expected splitting from
the magnetic field and the results are shown in Fig.~8.
Note that even after averaging a clear signature of the magnetic field is seen
in the splitting coefficients. 
Since we are comparing the average over the same set of modes for
the observed splittings and computed splittings for magnetic
field, we should be able to get some estimate of magnetic field if
a strong enough field does indeed exist. From Fig.~8 it can be seen that
there is no clear signature of any feature
near the base of the convection zone in the observed splittings,
and hence we can only set an
upper limit on the magnetic field in this layer. This will of course,
depend on the thickness of the magnetic layer.
Since there is no clear signature of any signal near the base of the
convection zone, for quantitative purpose we take the difference
between the lowest and highest point in the range $0.6<r_t/R_\odot<0.8$
in observed splitting coefficients. For $a_2$ this difference is
8.7 nHz for MDI and 7.0 nHz for GONG data, while computed splittings
with $\beta_0=10^{-4}$ show a difference of 12.6 nHz for a half-thickness
of $0.02R_\odot$. Thus, we can
put an upper limit of $0.7\times10^{-4}$ on $\beta_0$ which corresponds
to a magnetic field strength of 300 kG for a layer of 
half-thickness $0.02R_\odot$ near the base of the convection
zone. Similar analysis for splitting coefficient $a_4$ yields a
slightly larger upper limit of 400 kG.
These limiting values are close to what was obtained by Basu~(1997)
using a similar technique and is also consistent with the value
independently inferred by D'Silva \& Choudhuri (1993). Note,
this limit
roughly increases as $1/\sqrt{d}$, and clearly, if the thickness of this
region is smaller, the upper limit would be larger. It should be noted
that the tachocline, where the rotation rate undergoes a transition from
differential rotation in the convection zone to a solid-body like
rotation in the radiative interior may have a thickness as small as
$0.01R_\odot$ (Basu 1997; Antia, Basu \& Chitre 1998).
With this thickness the upper limit on magnetic field would
naturally be increased.

\begin{figure*}
\hbox to \hsize{\resizebox{\figwidth}{!}{\includegraphics{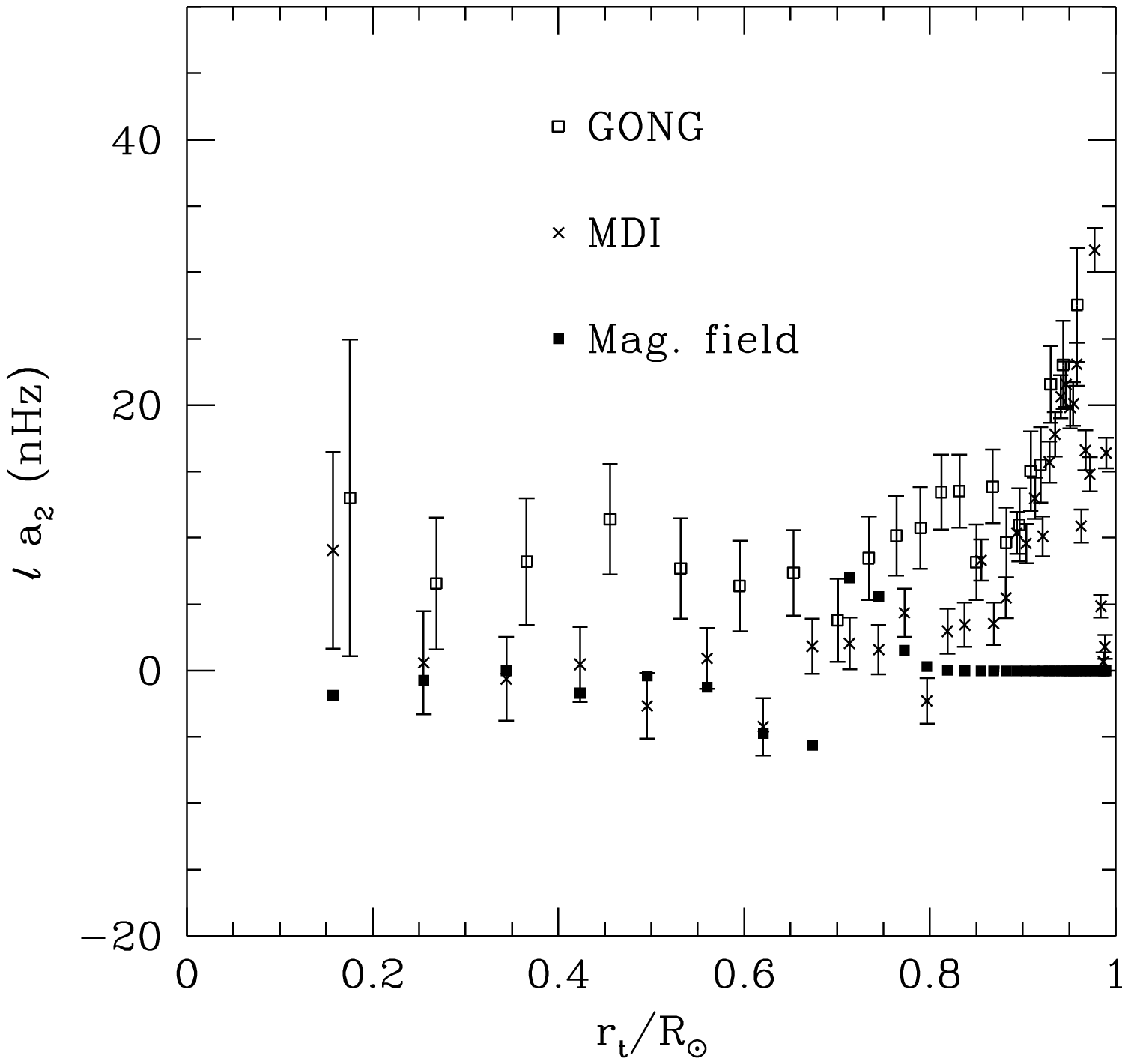}}
\hfil\resizebox{\figwidth}{!}{\includegraphics{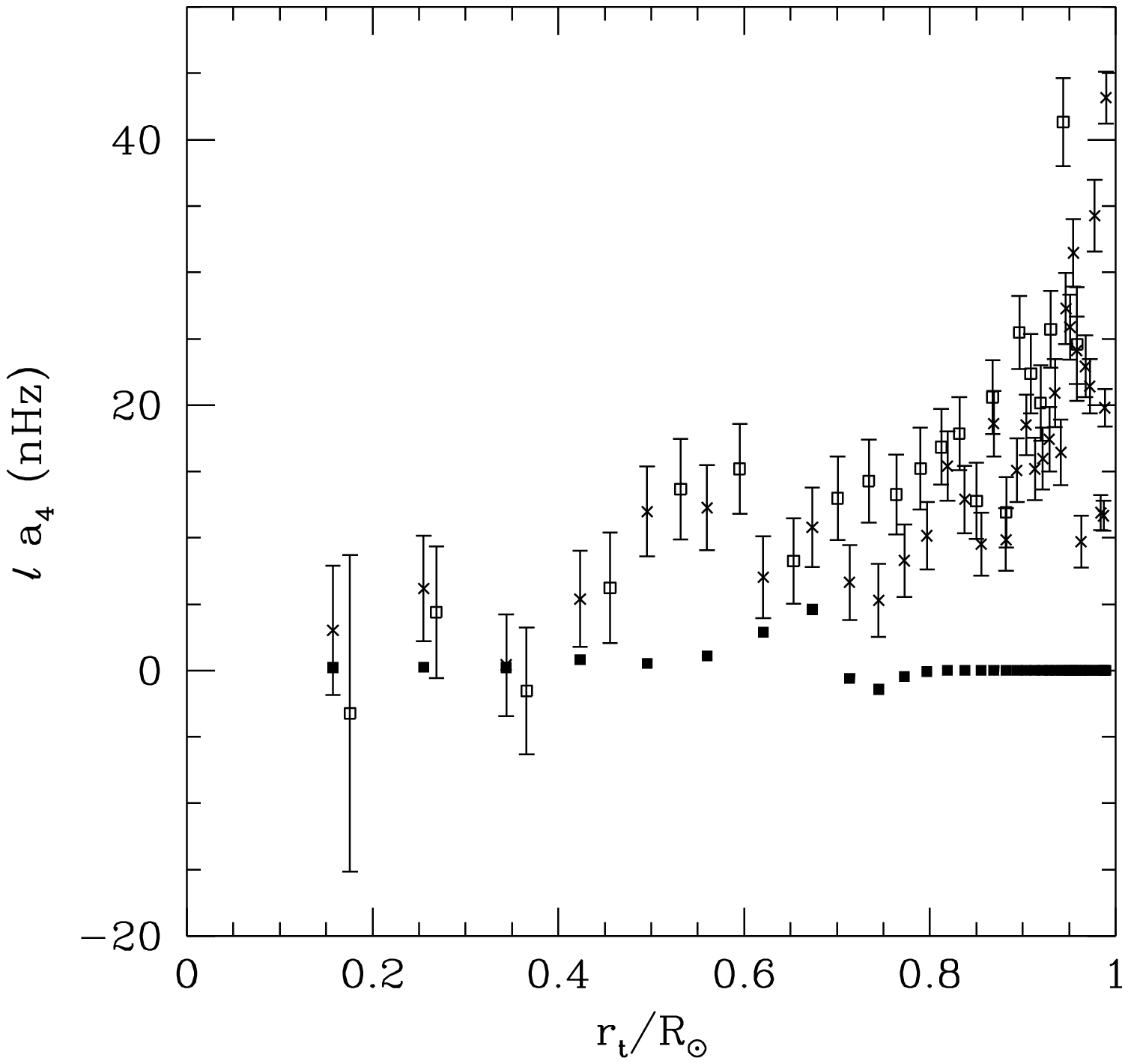}}}
\caption{
The splitting coefficients $a_2$ and $a_4$ from
a toroidal magnetic field concentrated near the base of the
convection zone, plotted
as a function of the lower turning point of the modes.
Each point represents an average over 30 neighbouring modes.
The estimated contribution from rotation has been subtracted
from the observed splittings plotted in the figure.
Magnetic field is given by Eqs.~(16,27) with $k=2$, $\beta_0=10^{-4}$,
$r_0=0.713R_\odot$ and $d=0.02R_\odot$.}
\end{figure*}

\begin{figure*}
\hbox to \hsize{\resizebox{\figwidth}{!}{\includegraphics{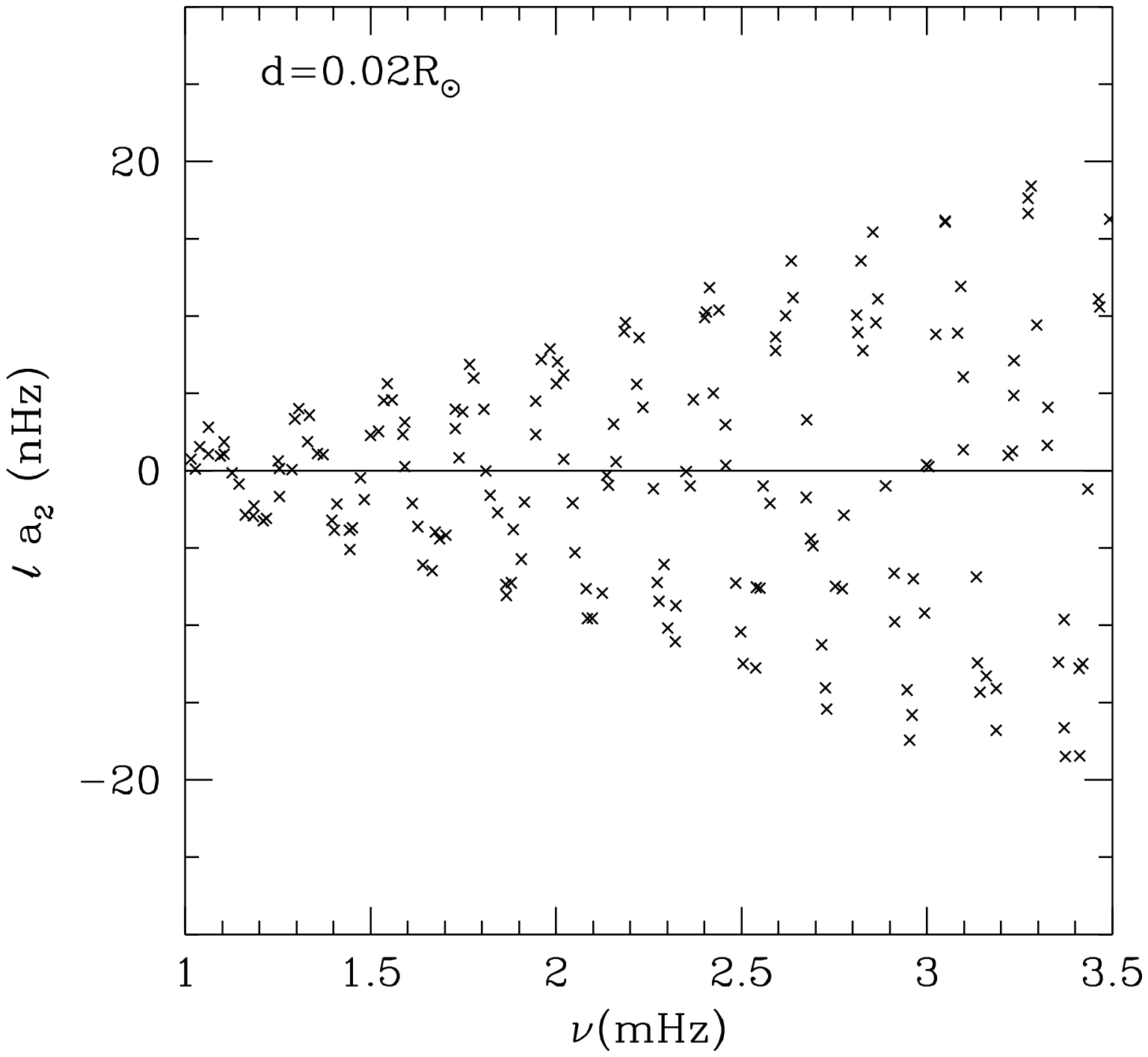}}
\hfil\resizebox{\figwidth}{!}{\includegraphics{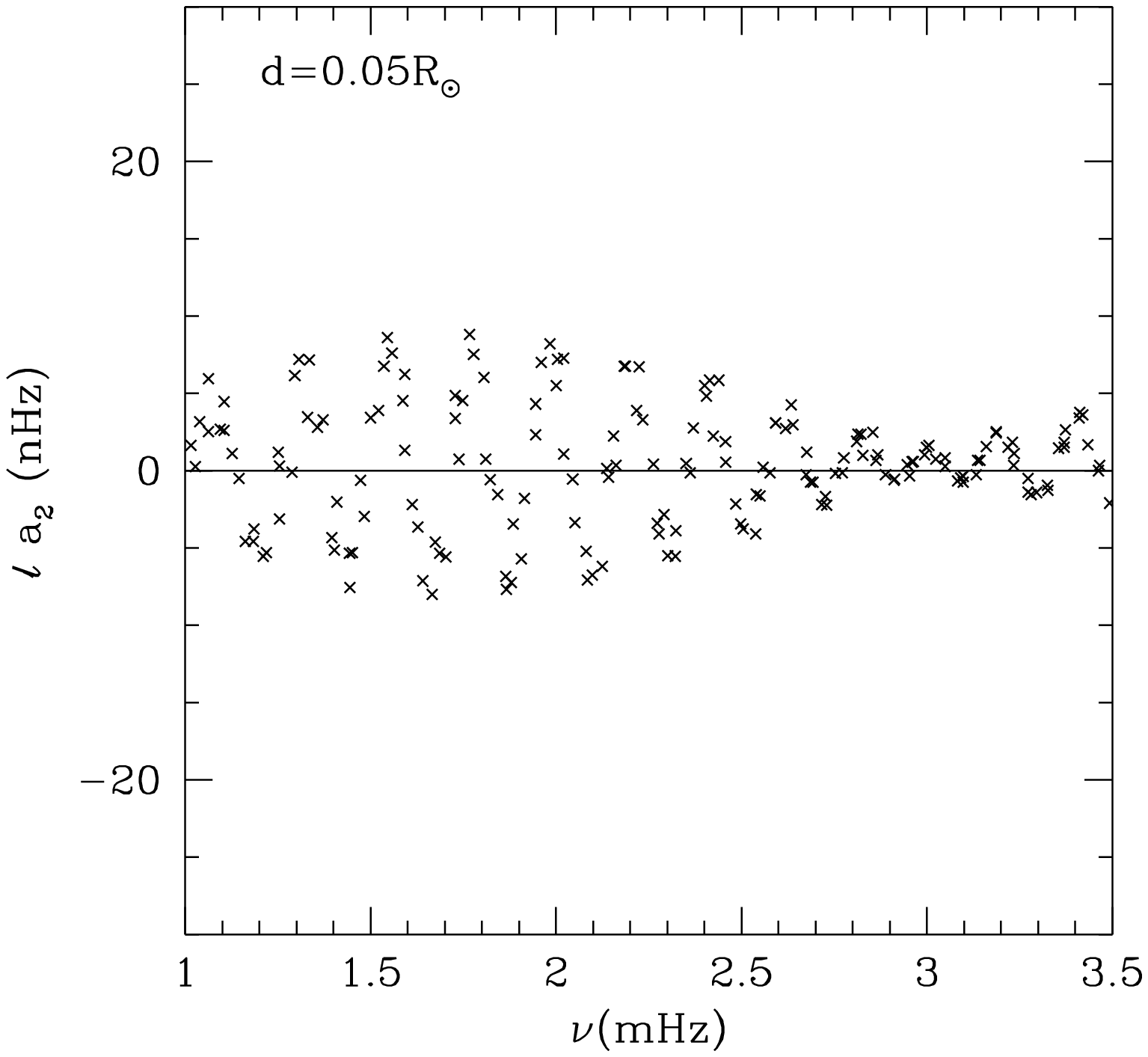}}}
\caption{
The splitting coefficient $a_2$ from
a toroidal magnetic field concentrated near the base of the
convection zone, plotted
as a function of the frequency for modes with $\ell\le10$.
Magnetic field is given by Eqs.~(16,27) with $k=2$, $\beta_0=10^{-4}$,
$r_0=0.713R_\odot$ and the value of $d$ as marked in the panels.}
\end{figure*}

\begin{figure*}
\hbox to \hsize{\resizebox{\figwidth}{!}{\includegraphics{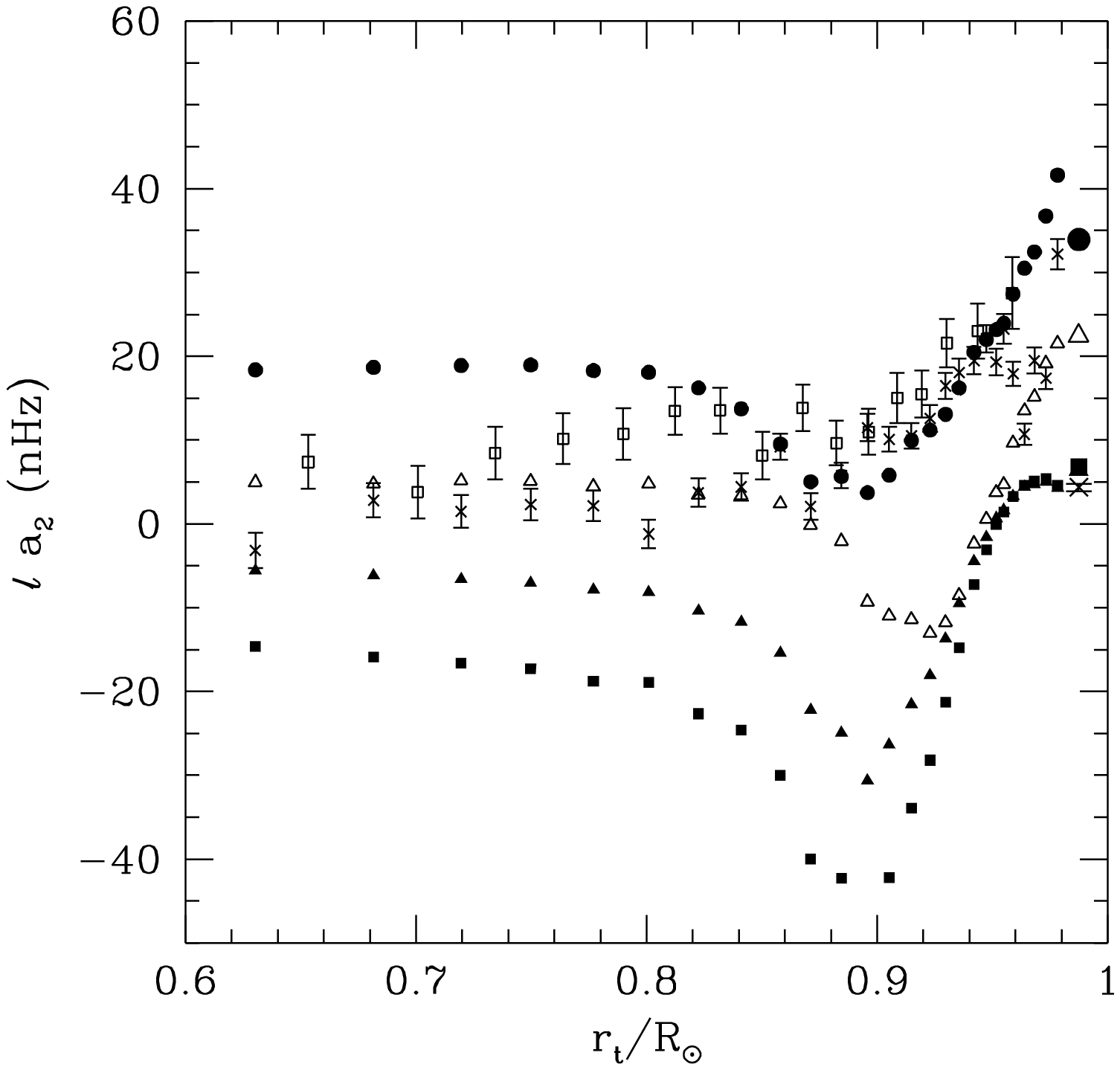}}
\hfil\resizebox{\figwidth}{!}{\includegraphics{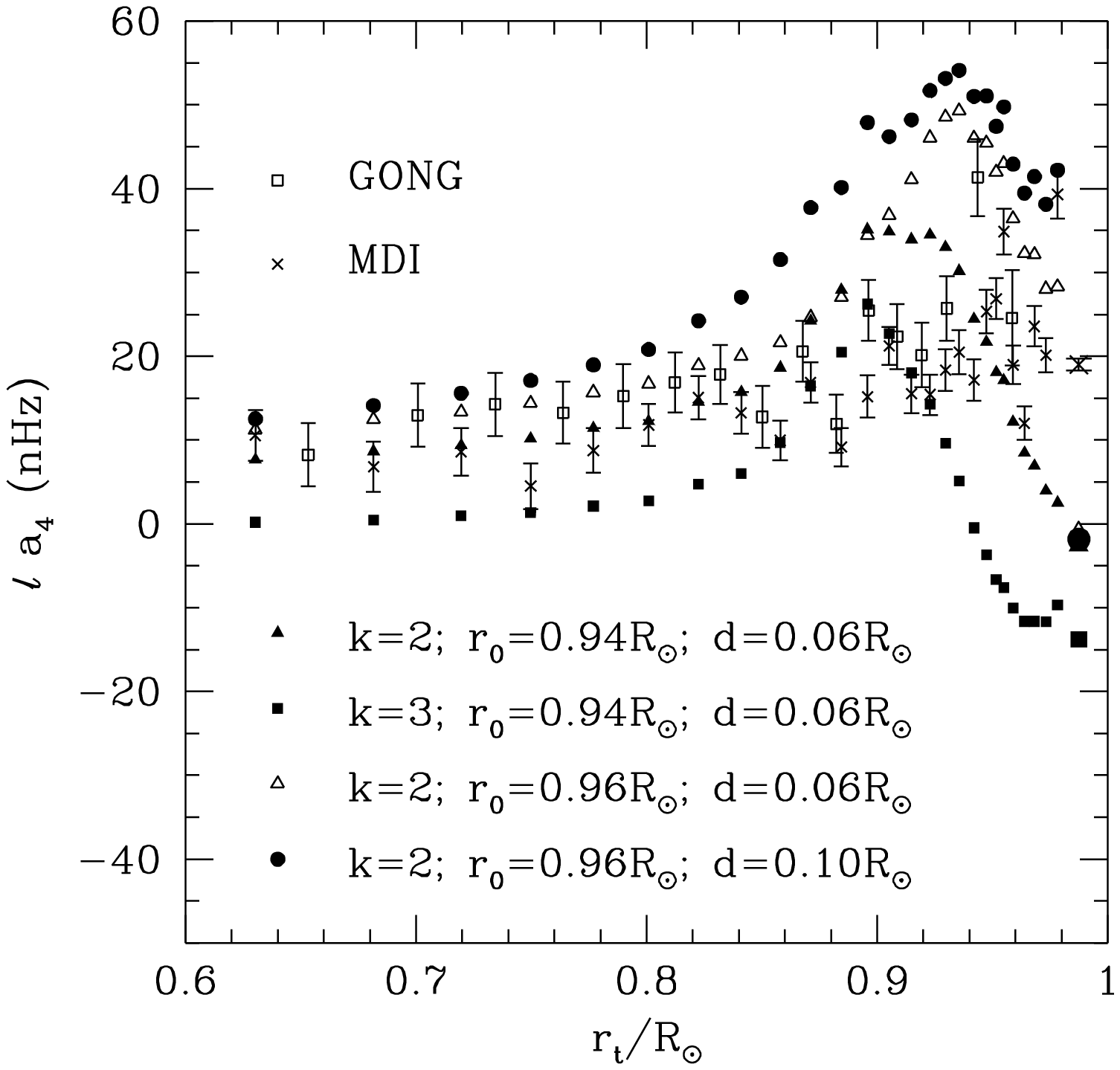}}}
\caption{
The splitting coefficients $a_2$ and $a_4$ from
a toroidal magnetic field concentrated in the upper part of the
convection zone, plotted
as a function of the lower turning point. The larger symbol near the
surface represents the average over all f-modes.
The estimated contribution from rotation has been subtracted
from the observed splittings plotted in the figure.
Magnetic field is given by Eq.~\ref{magtor} with $\beta_0=10^{-4}$,
and the value of $r_0,d$ and $k$ as marked in the figure.}
\end{figure*}

There is the possibility of distinguishing
seismologically between magnetic layers of different thicknesses
by using modes that penetrate well beneath the magnetic layer. A thin
layer will induce a signature in the $a_2$ and $a_4$ coefficients
which is periodic in mode frequency
(Gough \& Thompson 1988; Vorontsov 1988; Thompson 1988),
in much the same way that
the rather sharp transition near the base of the convective envelope
produces a periodic
signature in the mean frequencies (e.g., Gough 1990; Basu, Antia
\& Narasimha 1994; Monteiro, \jcd\ \& Thompson 1994).
Indeed it is this signature which is largely
responsible for the vertical spread of points for modes with
turning points at radii $r\la 0.6R$ in Fig.~7.
Basu~(1997) attempted to use this
oscillatory signal to obtain an
upper limit on magnetic field near the base of convection zone
(see also Gough \& Thompson~ 1988).
Fig.~9 shows $\ell a_2$ for modes with $\ell \le10$ for magnetic
field concentrated near the base of the convection zone, with
two different values of $d$.
It is clear that the amplitude of oscillatory signal varies significantly
with $d$. However, the observed splitting coefficients for low values
of $\ell$ have large errors and it is difficult to extract the
small oscillatory signal from these.

\subsection{Field in the upper convection zone}

Having considered a magnetic 
field at the base of the convection zone, where theory
suggests a field might be stored, we consider where else the data might
indicate the presence of magnetic field.
There is no signature for the presence of significant magnetic field in the
radiative interior, since the averaged residual splitting after
correcting for rotation seem to
be consistent with
zero. However, within the convection zone there is some significant
residual splitting, which could be due to the effect of a magnetic field.
An inspection of these residuals indicates the existence of a peak
around
$r=0.96R_\odot$, and indeed, if it is due solely to magnetic field,
the field may be
distributed around this depth ($\approx 28000$ km). It may be noted
that this is approximately
the depth to which shear layer seen in rotation profile extends
(Antia, Basu \& Chitre 1998; Schou et al.~1998).

We now attempt to estimate splittings due
to the field concentrated in this region.
Fig.~10 shows the splittings due to a few magnetic field configurations
which are concentrated in the upper part of the convection zone.
A comparison of these with the observed splittings indicates that there
may be an azimuthal magnetic field with $\beta<10^{-4}$ (i.e., $B\approx
20000$ G), with peak around $r=0.96R_\odot$. 

The possible existence of a magnetized layer with field of order 20 kG
located around $r=0.96R_\odot$ is, indeed, a significant inference
drawn from our analysis. The physical interpretation for the origin
of such a moderately strong magnetic field at this depth
below the Sun's surface is naturally a challenging task for theories
of solar dynamo to accommodate. It may be useful to recall here that
the numerical simulations of the Sun's outer convection zone (Nordlund
1999) indicate a major presence of downward moving plumes.
It is conceivable that these downdrafts could gather the turbulent
magnetic field in the sub-surface layers and carry them to depths
in the convective envelope until some sort of equipartition is
reached. Interestingly, the density, $\rho$ at a depth of 25--30 Mm
is upwards of $4\times10^{-3}$ gm cm$^{-3}$, while the downward velocity
for the plumes is of order 500 m s$^{-1}$. The dynamical pressure of the
plumes, $\rho v^2\ga 10^7$ dyne cm$^{-2}$, then becomes comparable
with the magnetic pressure, $B^2/8\pi$, corresponding to a field
strength of 20--30 kG. It is, therefore, tempting to envisage the
formation of such a magnetized layer by the pounding of the
downdrafts which tend to concentrate the field at depths where
the equipartition of the kind outlined above is approached.

In this study we have assumed a smooth toroidal magnetic field, but in
practice we do not expect such a field inside the convection zone.
Turbulence may be expected to randomize the magnetic field and such
a field may not be expected to produce any significant distortion in
the equilibrium state. The direct effect of magnetic field will still
be felt though the contribution would be different. Thus our results may be
treated as indicating the order of magnitude of field that may be
expected if the observed splitting coefficients are indeed due to
the magnetic field. If the field is concentrated in flux tubes which
occupy only a small fraction of the volume then the required magnetic
field could be correspondingly larger. If we assume that the flux tubes
occupy a fraction $f$ of the total volume, the magnetic field strength should
increase by $1/\sqrt{f}$.  If we consider only direct contribution to the
splittings then it turns out that $a_2$ is always negative for all
toroidal field configurations that we tried and hence such a contribution is
not likely to explain the observed splittings. But a different
magnetic field configuration, e.g., poloidal field might produce
$a_2$ with the required sign using only direct contribution.
The order-of-magnitude splitting caused by a magnetic field is
$\ell a_2/\nu\sim\beta_0\sim v_A^2/c_s^2$, where $v_A$ is the
Alfv\'en speed.
We therefore regard it as unlikely that a different magnetic field
configuration would produce a markedly different
answer for the field strength required to account for the
observed signal in $a_2$ and $a_4$.

A nonmagnetic latitudinally-dependent
perturbation to the wave propagation speed might be responsible for the signal
we detect (cf., Gough \& Zweibel 1995).
Once again we may expect a perturbation of order $10^{-4}$
located in the region around $r=0.96R_\odot$ to yield the observed
splittings.
Gough et al.~(1996) inferred a perturbation of
that magnitude, of unspecified origin, from earlier GONG data.
A temperature variation of order 10K,
suitably confined, might conceivably produce a similar signature.
In fact, Kuhn (e.g., Kuhn 1996) has argued that the thermal shadow of
belts of magnetic flux near the bottom of the convection zone can have
a significant effect on the even a-coefficients.  But Kuhn's models
show the largest temperature perturbations
occurring in the very superficial superadiabatic layer, at a depth of
a small fraction of one per cent of the solar radius. Such a perturbation
alone would be consistent with the f-modes having a small residual splitting,
but would not explain the apparent overturning of the p-mode splittings
at $r_t\approx 0.96R$. A magnetic field at some depth below the surface
may explain both aspects.
We certainly do not rule out the possibility that some
nonmagnetic asphericity, which we have not considered in detail
in this study, may account for some of the observed splittings.

\section{Conclusions}

Second order correction to mean frequencies due
to rotation is comparable to the error estimates in the observed
frequencies.
The error in helioseismic inversion introduced by
the frequency shift due to rotation is $\la 10^{-4}$, which is much
smaller than the estimated errors in inversions.
Further, a part of this frequency shift is expected to be nullified by the
general relativistic effects.
The shift in f-mode frequencies due to rotation
can reduce the estimated solar radius by 4 km.
The distortion due to rotation can yield surface oblateness of
$-5.8\times10^{-6}$ and $6.2\times10^{-7}$ in the $P_2(\cos\theta)$
and $P_4(\cos\theta)$
components, respectively. This is in reasonable agreement with
observed oblateness at the solar surface (Kuhn et al.~1998) and
it appears that most of the observed distortion is accounted by
the seismically inferred rotation rate in solar interior.
The quadrupole moment
$J_2=-2.18\times10^{-7}$ resulting from rotational distortion is
small enough to maintain consistency of the general theory of
relativity.

After subtracting the estimated contribution from
rotation to the splitting coefficients $a_2$ and $a_4$ from the
observed splittings, there is a small residual which is statistically
significant in the convection zone. This could arise from
a magnetic field.  From the magnitude of residual in observed splittings
we can tentatively conclude that magnetic field with
$\beta\approx10^{-4}$ may be present in
the upper part of the convection zone.
This corresponds to an azimuthal magnetic field of $\approx20$ kG around
$r=0.96R_\odot$. However, we cannot rule out the possibility that
this signal in splitting coefficients may arise from some
aspherical perturbation to the temperature field. This would be
practically indistinguishable from the effect of a magnetic field
using just the mode frequencies (Gough \& Zweibel 1995); but
complementary analyses such as time-distance helioseismology
might be able to distinguish them, since the local direct effect
of a magnetic field on the waves is anisotropic, whereas that of a
temperature perturbation is not.

A toroidal magnetic field that is concentrated near
the base of the convection zone gives a characteristic pattern in
the splittings for modes with lower turning point in that region.
Since no such signal is seen in observed frequencies, we can put an
upper limit of about 300 kG on the strength of the magnetic field
in this region.

\begin{acknowledgements}
We thank J.-P. Zahn for useful comments,
and J. Schou for providing MDI splittings. 
This work utilizes data obtained by the Global Oscillation Network
Group (GONG) project, managed by the National Solar Observatory, a
Division of the National Optical Astronomy Observatories, which is
operated by AURA, Inc. under a cooperative agreement with the
National Science Foundation.
The data were acquired by instruments operated by the Big
Bear Solar Observatory, High Altitude Observatory,
Learmonth Solar Observatory, Udaipur Solar Observatory,
Instituto de Astrof\'{\i}sico de Canarias, and Cerro Tololo
Interamerican Observatory.
This work also utilizes data from the Solar Oscillations
Investigation / Michelson Doppler Imager (SOI/MDI) on the Solar
and Heliospheric Observatory (SOHO).  SOHO is a project of
international cooperation between ESA and NASA. MJT acknowledges 
the support of the UK Particle Physics and Astronomy Research Council.
\end{acknowledgements}

\end{document}